\documentclass[aps,twocolumn,reprint,superscriptaddress,amsmath,amssymb]{revtex4-1}

\usepackage{graphicx}
\usepackage{dcolumn}
\usepackage{bm}
\usepackage{color}

\begin{document}


\title{$A$-cation control of magnetoelectric quadrupole order  in
$A$(TiO)Cu$_4$(PO$_4$)$_4$ ($A$ = Ba, Sr, and Pb)}

\author{K. Kimura}
\email[]{kentakimura@edu.k.u-tokyo.ac.jp}
\affiliation{Department of Advanced Materials Science, University of Tokyo, Kashiwa, Chiba 277-8561, Japan}

\author{M. Toyoda}
\affiliation{Department of Physics, Tokyo Institute of Technology, Meguro-ku, Tokyo 152-8550, Japan}
\author{P. Babkevich}
\affiliation{Laboratory for Quantum Magnetism, Institute of Physics, \'Ecole Polytechnique F\'ed\'erale de Lausanne (EPFL), CH-1015 Lausanne, Switzerland}
\author{K. Yamauchi}
\affiliation{ISIR-SANKEN, Osaka University, Ibaraki, Osaka 567-0047, Japan}
\author{M. Sera}
\affiliation{Division of Materials Physics, Graduate School of Engineering Science, Osaka University, Toyonaka, Osaka 560-8531, Japan}
\author{V. Nassif}
\affiliation{Univ. Grenoble Alpes, Inst NEEL, F-38000 Grenoble, France}
\affiliation{CNRS, Inst NEEL, F-38000 Grenoble, France}
\author{H. M. R\o{}nnow}
\affiliation{Laboratory for Quantum Magnetism, Institute of Physics, \'Ecole Polytechnique F\'ed\'erale de Lausanne (EPFL), CH-1015 Lausanne, Switzerland}
\author{T. Kimura}
\affiliation{Department of Advanced Materials Science, University of Tokyo, Kashiwa, Chiba 277-8561, Japan}

\email[*]{kentakimura@edu.k.u-tokyo.ac.jp}
\altaffiliation{aaa}


\date{\today}

\begin{abstract}
Ferroic magnetic quadrupole order exhibiting macroscopic magnetoelectric activity is discovered in the novel compound $A$(TiO)Cu$_4$(PO$_4$)$_4$ with $A$ = Pb, which is in contrast with antiferroic quadrupole order observed in the isostructural compounds with $A$ = Ba and Sr.
Unlike the famous lone-pair stereochemical activity which often triggers ferroelectricity as in PbTiO$_3$, the Pb$^{2+}$ cation in Pb(TiO)Cu$_4$(PO$_4$)$_4$ is stereochemically inactive but dramatically alters specific magnetic interactions and consequently switches the quadrupole order from antiferroic to ferroic.
Our first-principles calculations uncover a positive correlation between the degree of $A$-O bond covalency and a stability of the ferroic quadrupole order.

\end{abstract}



\maketitle


\section{INTORODUCTION}
Earlier works demonstrated that the usage of specific elements with characteristic chemical properties is effective to realize desired ferroic order. For example, lone-pair stereochemical activity of a heavier post-transition metal cation with an $s^{2}$ electron configuration such as Pb$^{2+}$ and Bi$^{3+}$, which we call an $s^{2}$-cation, is known as a driving force for ferroelectric order \cite{Shimoni1998,Walsh2011}, as discussed in perovskite oxides PbTiO$_3$ \cite{Cohen1992}, BiMnO$_3$ \cite{Seshadri2001}, and BiFeO$_3$ \cite{Ravindran2006}. The stereochemically active $s^2$-cations are surrounded by ``hemidirected'' local coordination, in which there is a void in the distribution of bonds to the ligands \cite{Shimoni1998}.
The origin for this directional bonding is explained by the hybridization between nominally empty metal $p$ states with anti-bonding states formed by filled metal $s$ states and ligand $p$ states \cite{Walsh2011}. Such a hybridization is possible only when the inversion-symmetry at the cation site is broken. This is a driving force for off-center distortion and thus ferroelectric order.

Another potential role of $s^2$-cations is an impact on magnetism in insulating magnetic oxides. There, dominant magnetic superexchange interactions are usually mediated by O 2$p$ orbitals near Fermi energy ($E_{\rm F}$) \cite{Anderson1950}.
As exemplified by the comparison between PbTiO$_3$ and BaTiO$_3$ \cite{Cohen1992}, $s^2$-cations tend to exhibit strong orbital hybridization with O ions,
which should significantly affects superexchange interactions. Note that $s^2$-cations are not necessarily stereochemically active; there are comparable number of compounds containing such cations located in ``holodirected'' local environment without a void in the ligand bond distribution \cite{Shimoni1998}. In this case, substituting $s^{2}$-cations can be a promising way of fine-tuning magnetic interactions without large distortion of the original structure.

Among various ferroic orders, a particular class of magnetic order with broken space-inversion and time-reversal symmetries has recently attracted considerable interest because it can exhibit symmetry-dependent unique phenomena, such as magnetoelectric (ME) effects  \cite{Astrov1960,Dubovik1990,Schmid2001,Fiebig2005,VanAken2007,Spaldin2008,Spaldin2013,Yamaguchi2013,Zimmermann2014} and unconventional nonreciprocal electromagnetic responses \cite{Barron2004,Arima2008,Bordacs2012}.
From a symmetry point of view, it is known that ferroic order of magnetic multipole moments (toroidal, monopole, and quadrupole moments) fulfills this symmetry condition \cite{Dubovik1990,Schmid2001,Spaldin2008,Spaldin2013}.
Thus far, several materials have been reported to exhibit ferroic order of toroidal \cite{VanAken2007,Zimmermann2014,Toledano2015} and quadrupole moments \cite{Arima2008,staub2009parity}.
However, the experimental realization of ferroic order composed of a single multipole component (i.e., ferroic order of pure toroidal, monopole, or quadrupole moments) is extremely rare.
Here, we employ the above-mentioned idea ---$s^2$-cation control of magnetism--- to design and realize pure ferroic magnetic quadrupole order with macroscopic ME activity.

\begin{figure*}[t]
\includegraphics[width=16cm]{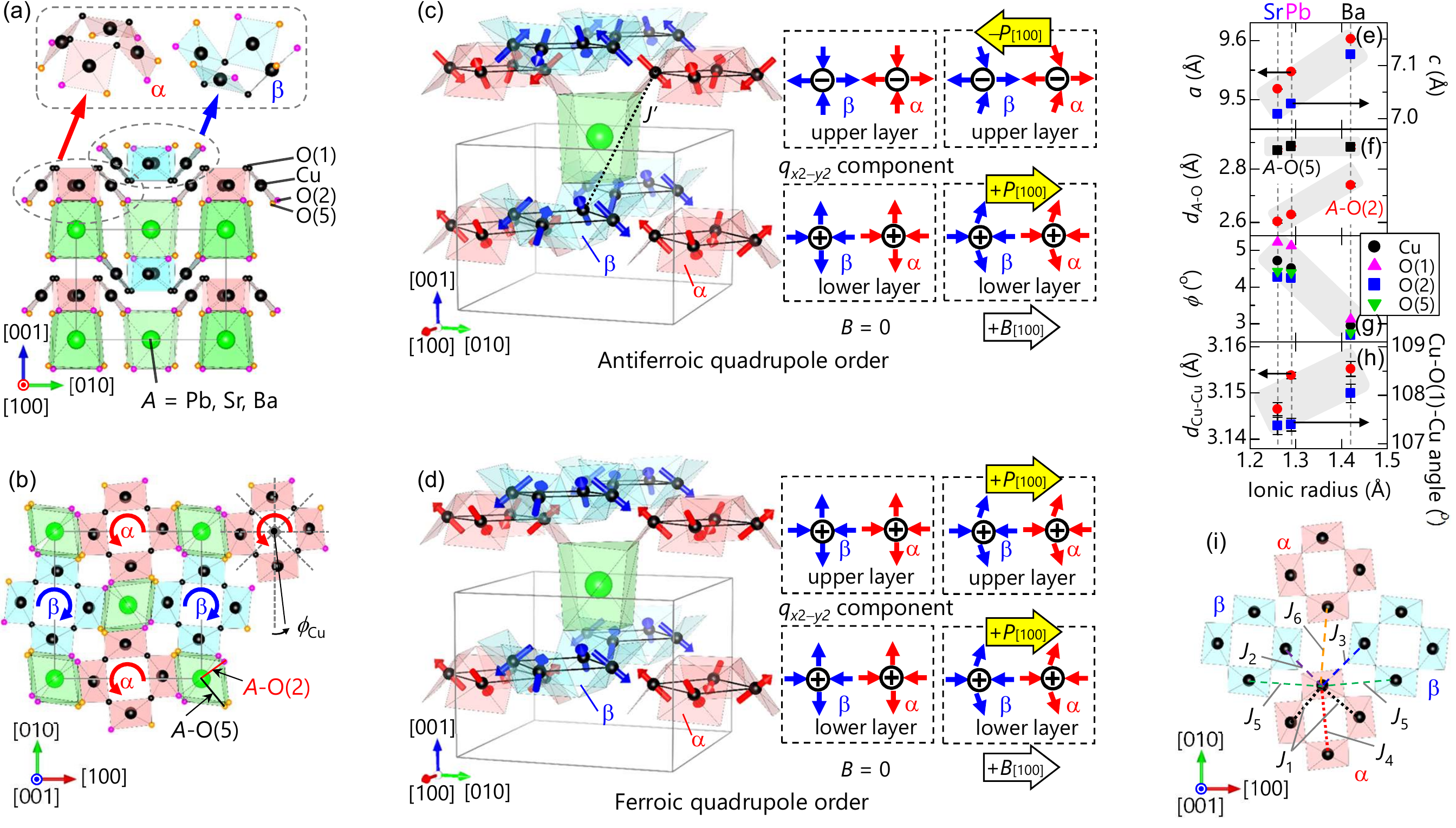}
\caption{
Description of the crystal and magnetic structures of the $A$(TiO)Cu$_4$(PO$_4$)$_4$ system.
(a) The crystal structure viewed along the [100] direction. The grey line represents a unit cell. Two types of Cu$_4$O$_{12}$ square cupola clusters, $\alpha$ (upward) and $\beta$ (downward) are emphasized. 
(b) The top view of the crystal structure. The thick red and blue arrows for cluster $\alpha$ and $\beta$ represent the rotation direction upon the antiferrorotative distortion. The rotation angle $\phi$ for Cu is depicted as an example.
(c) Antiferroic and (d) ferroic magnetic quadrupole orders. The interlayer exchange coupling $J'$ mediated by the $A$O$_8$ polyhedron is indicated in (c). The $q_{x^{2}-y^{2}}$ components of each layer at zero magnetic field ($\bf B$) and their signs are illustrated. Electric polarization ($\bf P$) induced by $\bf B$ parallel to the [100] direction is also shown.
(e)-(h) $A$-cation ionic radius dependence of lattice constants $a$ and $c$, $A$-O bond length $d_{A-\rm {O}}$, rotation angle $\phi$, and Cu-Cu bond length $d_{\rm {Cu-Cu}}$ and Cu-O(1)-Cu bond angle. The gray lines represent trends of the ionic radius dependence on the respective parameters.
(i) Definition of intralayer exchange couplings $J_k$ ($k = 1-6$).
\label{fig1}
}
\end{figure*}

We focus on oxide magnetic insulators $A$(TiO)Cu$_4$(PO$_4$)$_4$; the $A$ = Ba and Sr systems (abbreviated as BaTCPO and SrTCPO, respectively) were recently synthesized \cite{KKimura2016}. They are isostructural and crystallize into the chiral tetragonal structure with space group $P42_{1}2$. Our previous reports on BaTCPO revealed that the magnetic properties are dominated by layered arrangement of unique Cu$_4$O$_{12}$ square-cupola clusters in the $ab$-plane [Fig.~\ref{fig1}(a)], which is characterized by an antiferrorotative distortion of two types of Cu$_4$O$_{12}$ labeled $\alpha$ and $\beta$ with a rotation angle $\phi$ [Fig.~\ref{fig1}(b)] \cite{KKimura2016b,Kato2017}. The magnetic layers are sandwiched by nonmagnetic layers containing $A^{2+}$ cations that govern quasi two-dimensional (2D) magnetism with weak interlayer couplings [Fig.~\ref{fig1}(a)].
BaTCPO undergoes an antiferromagnetic (AFM) phase transition at $T_{\rm N}^{\rm Ba} \approx ~ 9.5$ K. In the magnetic structure below $T_{\rm N}^{\rm Ba}$ [Fig.~\ref{fig1}(c)], the in-plane components of magnetic moments on each Cu$_4$O$_{12}$ cluster form a local quadrupole moment defined as $q_{ij}\propto \Sigma_{n}(r_{ni} S_{nj} + r_{nj} S_{ni} - \frac{2}{3} \delta_{ij} {\bf r}_{n} \cdot {\bf S}_{n})$ \cite{Dubovik1990,Schmid2001,Spaldin2008,Spaldin2013}. Here, $n$ represents a label of the spin ${\bf S}_{n}$ at the position vector ${\bf r}_{n}$ and $i$, $j$ denote the $x$, $y$, or $z$ axis parallel to the [100], [100], or [001] axis, respectively.
Note that this magnetic quadrupole moment, defined by multi-site spins, is different from an atomic magnetic quadrupole moment which has been recently proposed to be an order parameter for the pseudo-gap phase of cuprate superconductors \cite{lovesey2015ferro,fechner2016quasistatic}.
The quadrupole moments of BaTCPO are mostly composed of $q_{x^{2}-y^{2}}$ components. The $q_{x^{2}-y^{2}}$ components on different Cu$_4$O$_{12}$ clusters align uniformly in the $ab$-plane, but in the antiparallel fashion along the $c$-axis, which means a layered antiferroic quadrupole order. The associated clear magnetodielectric signals have been experimentally observed \cite{KKimura2016b} and theoretically explained on the basis of an effective $S=1/2$ spin model with Dzyaloshinskii-Moriya interactions \cite{Kato2017}. However, the antiferroic nature of the magnetic quadrupole order prohibits macroscopic linear ME effects, namely, magnetic-field ($\bf B$) induced electric polarization ($\bf P$) or electric-field ($\bf E$) induced magnetization ($\bf M$), which hampers exploring macroscopic ME phenomena arising from magnetic quadrupoles.
Considering that the $A$O$_{8}$ polyhedra bridge the Cu$_4$O$_{12}$ clusters in the neighboring layers, it is expected that inserting $s^2$-cations into the $A$-site instead of Ba$^{2+}$ may tune interlayer magnetic couplings and thus induce ferroic quadrupole order as illustrated in Fig.~\ref{fig1}(d). 

In this paper, we choose Pb$^{2+}$ as the $s^{2}$-cation because it is isoelectric and has ionic radius (1.29 \AA) in-between Sr$^{2+}$ (1.26 \AA) and Ba$^{2+}$ (1.42 \AA) \cite{Shannon1976}. We have synthesized single crystals of Pb(TiO)Cu$_4$(PO$_4$)$_4$ (abbreviated as PbTCPO). Our experimental and theoretical studies on $A$TCPO ($A$ = Pb, Ba, and Sr) show that inserting Pb$^{2+}$ indeed switches the quadrupole order from antiferroic to ferroic.

\section{EXPERIMENTAL AND THEORETICAL METHODS}
Single crystals of BaTCPO and SrTCPO were grown by the flux method \cite{KKimura2016}, while those of PbTCPO were grown by the slow cooling from the melt of stoichiometric mixtures. Powder X-ray diffraction (XRD) measurements on crushed single crystals confirmed a single phase. The single crystal XRD measurement on PbTCPO was performed at room temperature on a Rigaku XtaLAB P200 diffractometer with confocal-monochromated Mo K$\alpha$ radiation ($\lambda = 0.71075$ \AA). A small single crystal of PbTCPO was used for the measurements. Data were collected and processed using CrystalClear-SM Expert 2.1 b45 software (Rigaku, 2015) to apply empirical absorption correction. The refinement of the crystal structure was performed using the CrystalStructure crystallographic software package (Rigaku).

For the measurements of bulk properties, the crystals were oriented using the Laue XRD method. Measurements of $\bf M$ down to temperature ($T$) of 1.8 K and $\bf B$ up to 7 T were performed using a commercial superconducting quantum interference device magnetometer (Quantum Design MPMS3). The specific heat ($C_P$) was measured down to 2 K by a thermal relaxation method using a commercial calorimeter (Quantum Design PPMS). For dielectric measurements, single crystals were cut into thin plates and subsequently the electrodes were made by painting a silver paste on the pair of widest surfaces. The dielectric constant ($\varepsilon$) was measured using an $LCR$ meter at an excitation frequency of 100 kHz. The $T$ or $\bf B$ dependence of $\bf P$ was obtained by integrating pyroelectric or magnetoelectric (ME) currents during warming or sweeping $\bf B$ measured using an electrometer (Keithley 6517). All the $\bf P$ measurements were performed at zero $\bf E$ after the sample was cooled down while applying $\bf B$ and $\bf E$ (ME cooling).

Neutron diffraction measurements on BaTCPO, SrTCPO, and PbTCPO were performed using the D20 diffractometer at Institut Laue-Langevin using a wavelength of 2.41 \AA. Powder samples of approximately 5g each were sealed in vanadium cans. Neutron diffraction powder patterns were collected between 2 and 17 K. To obtain high-statistics, powder patterns recorded in the magnetically ordered and paramagnetic phases were collected for 8 h at each temperature. A calibration using a standard Na$_2$Ca$_3$Al$_2$F$_{14}$ sample was performed in order to fix the instrumental resolution during the experiment. Refinement of the magnetic and nuclear structures were carried out using the Fullprof package \cite{fullprof}.

Density functional theory (DFT) calculations were performed to estimate the magnitude of dominant magnetic interactions. The VASP (Vienna ab initio simulation package) \cite{Kresse1993,Kresse1994,Kresse1996,Kresse1996_cms} was used with a projector-augmented wave basis set \cite{PAW,Kresse1999}. The exchange-correlation term was described by the Perdew-Burke-Ernzerhof generalized gradient approximation \cite{PBE1996}. The following states were treated as the valence states: 5$s$, 5$p$, and 6$s$ for Ba; 4$s$, 4$p$ and 5$s$ for Sr; 6$s$ and 6$p$ for Pb; 5$s$ and 5$p$ for Sn; 3$d$, 3$p$ 3$d$, and 4$s$ for Ti; 2$s$ and 2$p$ for O; 3$d$ and 4$s$ for Cu; and 3$s$ and 3$p$ for P. The plane-wave cut-off energy was $E_{\rm cut} = 500$ eV. The integration over Brillouin zone was performed by using a $4\times4\times5$ sampling mesh. The DFT+$U$ method \cite{Liechtenstein1995} was used for correction for strongly correlated Cu-3$d$ states. All the calculations were performed with $U_{\rm eff} = U - J = 4$ eV unless otherwise noted. With this value, the calculated Weiss temperature of BaTCPO is in good agreement with the experimental value \cite{KKimura2016b}. We have also confirmed that the different values of $U_{\rm eff}$ do not change our qualitative results and conclusion \cite{SMprb-PbTi}. In DFT+$U$ calculations for other Cu$^{2+}$ compounds such as superconducting cuprates La$_2$CuO$_4$ and Sr$_2$CuO$_2$F$_2$, $U_{\rm eff} = 5-7$ eV are commonly used \cite{morozov2015exchange}. Therefore, our value ($U_{\rm eff}=4$ eV) might be slightly too small for further quantitative discussion on transition temperatures and detailed electronic band structure, which, however, is not the scope of this paper. 

The magnetic exchange coupling constants are defined with an effective classical Heisenberg model 
\begin{eqnarray}
\begin{array}{c}
H = -\frac{1}{2} \displaystyle\sum_{l \ne m} J_{lm} {\bf e}_l \cdot {\bf e}_m.
\end{array}
\end{eqnarray}
Here, ${\bf e}_l$ is the unit vector pointing to the direction of the spin at site $l$, and $J_{lm}$ is the effective coupling constant between site $l$ and $m$. For intralayer couplings, we estimated the averaged coupling constants $J_k$ for each shell of $k$-th nearest neighbors (n.n.),
\begin{eqnarray}
\begin{array}{c}
J_k = 
\displaystyle\sum_{l \ne m}^{k-\rm{th~n.n.}} J_{lm} {\bf e}_l \cdot {\bf e}_m  \bigg/ \displaystyle\sum_{l \ne m}^{k-\rm{th~n.n.}} {\bf e}_l \cdot {\bf e}_m.%
\end{array}
\end{eqnarray}
Up to the sixth n.n. interactions ($k = 1-6$), $J_k$ were estimated as the best fit to the DFT total energies with different spin configurations. For the interlayer couplings, because they were found to be much smaller than the intralayer ones, we estimated a specific coupling $J'$ [Fig.~\ref{fig1}(c)] in a more direct method. We considered a fictitious material with the same crystal structure where the magnetic Cu$^{2+}$ ions are substituted by the nonmagnetic Zn$^{2+}$ ions except for the two specific sites that are coupled by $J'$. $J'$ is then given by
\begin{eqnarray}
\begin{array}{c}
2J' \approx E_{\uparrow \downarrow} - E_{\uparrow \uparrow},
\end{array}
\end{eqnarray}
where $E_{\uparrow \downarrow}$ and $E_{\uparrow \uparrow}$ are the DFT total energies of the fictitious material with antiparallel and parallel spin configurations, respectively, at the two Cu$^{2+}$ sites.

All the DFT calculations have been done with the experimental crystal structures at room temperature determined by single crystal XRD measurements, unless otherwise noted. 
To support this choice, we have performed the structural optimization for BaTCPO and confirmed that it does not change our conclusion qualitatively \cite{SMprb-PbTi}. Moreover, measurements of powder neutron diffraction at $T < 17$ K and the $T$-dependence of $M/B$ have detected no sign of a structural phase transition at low $T$s in all the three systems.

\section{RESULTS AND DISCUSSION}
Single-crystal XRD measurements reveal that PbTCPO is indeed isostructural with BaTCPO and SrTCPO \cite{SMprb-PbTi}. In Figs.~\ref{fig1}(e)-\ref{fig1}(h), we plot the main structural parameters as a function of ionic radii of $A$-cation ($r_{A}$).
The lattice constants solely increase as $r_{A}$ increases. Moreover, a change in the $A$-O(2) bond length causing a deformation of the $A$O$_{8}$ polyhedron is also monotonic, and the $A$ cation site symmetry (222) remains unchanged. 
These observations confirm that Pb cations in PbTCPO are stereochemically inactive.

\begin{figure}[t]
\includegraphics[width=8.6cm]{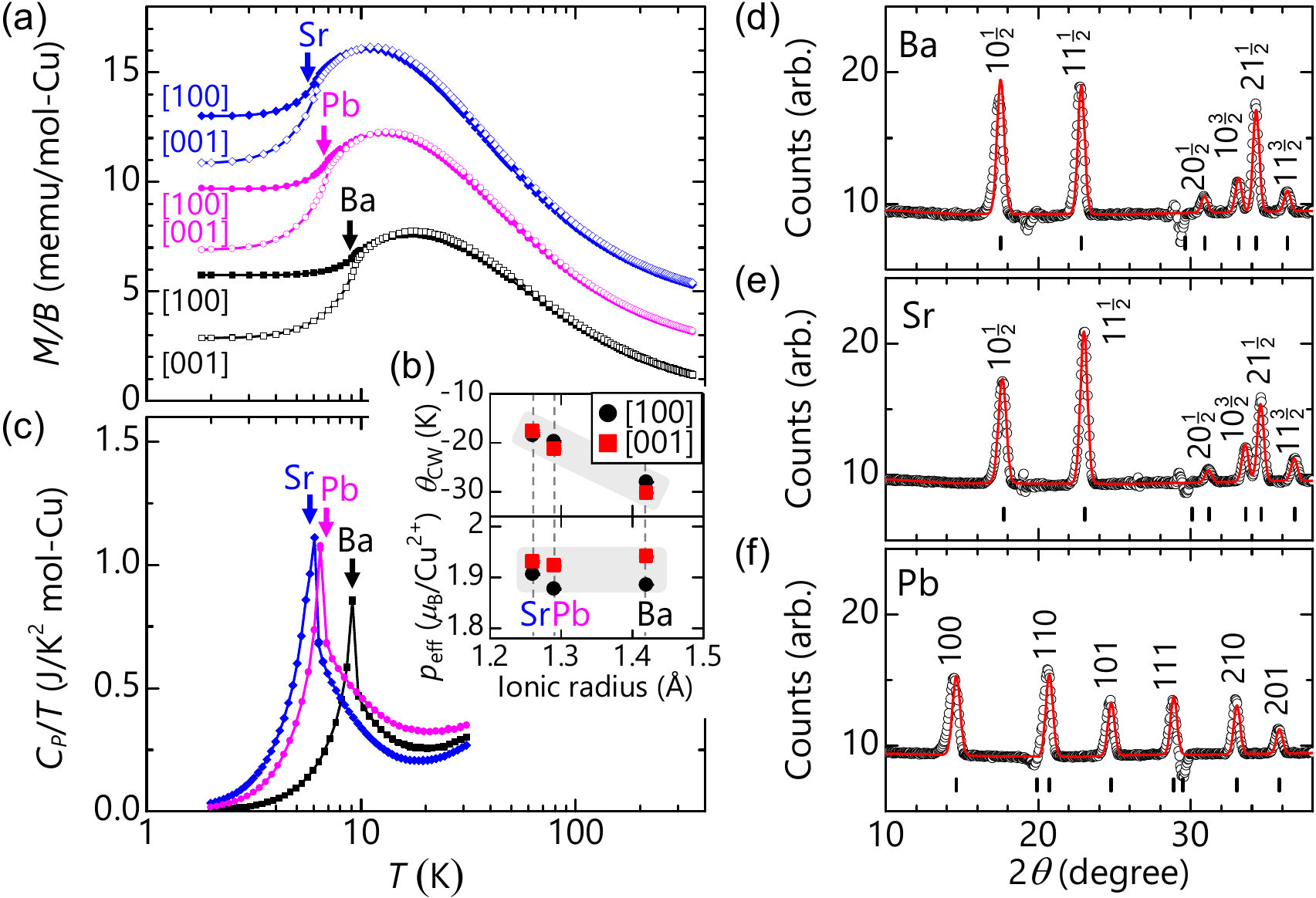}
\caption{
Magnetic properties and magnetic quadrupole order in the $A$(TiO)Cu$_4$(PO$_4$)$_4$ system.
(a) Temperature ($T$) dependence of magnetization ($\bf M$) divided by magnetic field ($\bf B$) applied along the [100] and [001] directions. The data for Pb and Sr are shifted by 1 and 2 memu/mol-Cu, respectively, for clarity. 
(b) $A$-site ionic radius dependence of Curie-Weiss temperature ($\theta_{\rm CW}$) and effective moment ($p_{\rm eff}$) obtained by fitting $M/B$ in (a) with the Curie-Weiss law. 
(c) Specific heat divided by $T$.
(d)-(f) Magnetic neutron diffraction patterns (open circles) in the antiferromagnetic phase for the Ba, Sr, and Pb systems after subtracting the structural contributions estimated in the paramagnetic phase. The solid red line shows the model fit discussed in the text. Vertical black lines indicate positions of magnetic reflections.
\label{fig3}
}
\end{figure}

To substantiate the above intuitive idea for realizing ferroic quadrupole order, we have measured the magnetic properties of PbTCPO, SrTCPO, and BaTCPO.
Figure \ref{fig3}(a) shows the temperature ($T$) dependence of $\bf M$ divided by $\bf B$ applied along the [100] and [001] axes for single crystals of the three systems. Fits to the high temperature $M/B$ data ($T > 100$ K) using the Curie-Weiss law reveal that the three systems have approximately the same effective moment around 1.9 $\mu_{\rm B}\rm{/Cu}$ [Fig.~\ref{fig3}(b)]. The Weiss temperatures  ($\theta_{\rm CW}$) are negative (AFM), whose magnitude decreases monotonically as $r_{A}$ decreases. This is reasonable because for smaller $r_{A}$ the Cu-O-Cu angle approaches to 90$^\circ$ [Fig.~\ref{fig1}(h)] which generally weakens Cu-O-Cu superexchange interactions. On cooling, $M/B$ shows an $A$-cation-independent feature characterized by a broad maximum followed by a clear anomaly at $T_{\rm N}$. The former suggests short-range correlations within each cluster and/or 2D layer and the latter indicates AFM long-range ordering due to weak interlayer couplings \cite{KKimura2016b}. These AFM transitions are also confirmed by a peak in specific heat $C_P$ [Fig.~\ref{fig3}(c)].
No metamagnetic transition is observed at $T=1.8$ K up to $|{\bf B}|=7$ T for the three systems \cite{SMprb-PbTi}.

Neutron scattering experiments on powder samples were employed for microscopic characterization of the magnetic properties. Figures \ref{fig3}(d)-\ref{fig3}(f) show neutron magnetic diffraction patterns in the AFM phase of the three compounds.
The diffraction pattern of SrTCPO is very similar to that of BaTCPO exhibiting antiferroic quadrupole order with $q_{x^{2}-y^{2}}$ components \cite{KKimura2016b}. The reflections can be indexed by a single propagation wavevector $\mathbf{k} = (0,0,0.5)$, which corresponds to a doubling of the unit cell along the $[001]$ direction. In addition, the best fit magnetic structure obtained by the Rietveld refinement \cite{fullprof} indicates that the spin arrangement of SrTCPO can also be characterized by the ordering of $q_{x^{2}-y^{2}}$ components on Cu$_4$O$_{12}$ \cite{SMprb-PbTi}. This result confirms that both BaTCPO and SrTCPO exhibit antiferroic magnetic quadrupole order as displayed in Fig.~\ref{fig1}(c).
For PbTCPO, a similar ordering of $q_{x^{2}-y^{2}}$ components appears. However, the magnetic unit cell is identical to the chemical one, as indicated by the magnetic reflections indexed by $\mathbf{k} = (0,0,0)$ \cite{SMprb-PbTi}. Therefore, inserting $s^{2}$-cations in this system indeed induces ferroic magnetic quadrupole order [Fig.~\ref{fig1}(d)].

\begin{figure}[t]
\includegraphics[width=8.6cm]{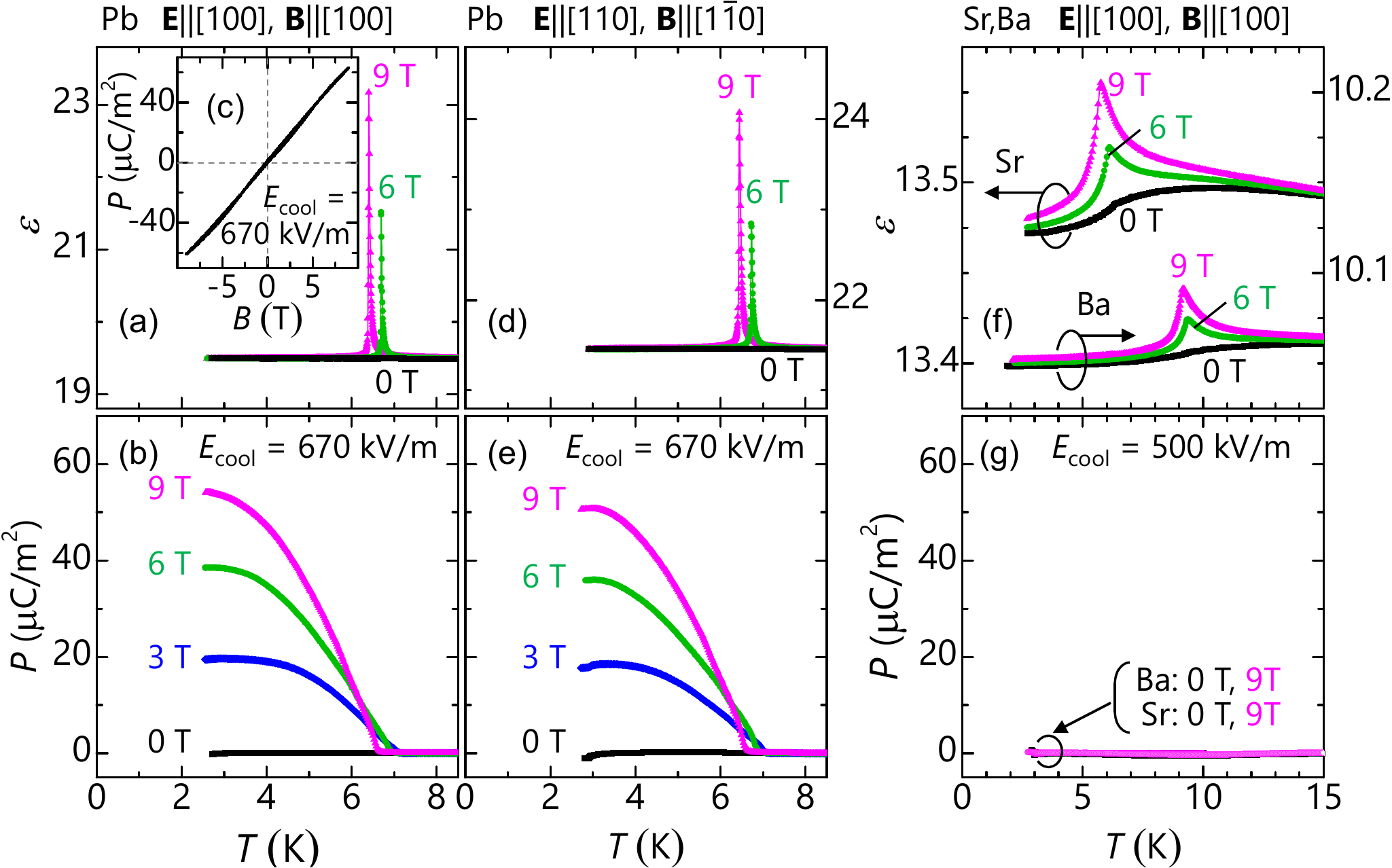}
\caption{
$A$-cation controlled magnetoelectric properties in the $A$(TiO)Cu$_4$(PO$_4$)$_4$ system.
Temperature ($T$) dependence of (a) the dielectric constant along the [100] direction $\varepsilon_{[100]}$ and (b) electric polarization $P_{[100]}$ for the Pb system in various magnetic fields applied along the [100] direction ($B_{[100]}$).
(c) Isothermal $P_{[100]}$-$B_{[100]}$ curve at 2.5 K.
$T$-dependence of (d) $\varepsilon_{[110]}$ and (e) $P_{[110]}$ in various strengths of $B_{[1\bar{1}0]}$ for the Pb system.
$T$-dependence of (f) $\varepsilon_{[100]}$  and (g) $P_{[100]}$ in various strengths of $B_{[110]}$ for the Ba and Sr systems. In (b), (c), (e), (g), the values of electric field applied during ME cooling ($E_{\rm cool}$) are indicated.
\label{fig4}
}
\end{figure}

Here, we provide experimental evidence for the ME activity arising from the ferroic quadrupole order in single crystalline specimens of PbTCPO. As discussed previously \cite{KKimura2016b}, the uniform ordering of $q_{x^{2}-y^{2}}$ components results in a magnetic point group of $4'22'$, which allows for a linear ME effect given by the ME tensor \cite{Birss1966},
\begin{eqnarray}
\alpha_{\rm ME} = \left(
    \begin{array}{ccc}
      \alpha_{xx} & 0 & 0 \\
      0 & -\alpha_{xx} & 0 \\
     0 & 0 & 0
    \end{array}
  \right).
  \label{eqn1}
\end{eqnarray}
This ME tensor predicts that $\bf P$ along the [100] direction ($ P_{[100]}$) is induced by applying $\bf B$ along the [100] direction ($B_{[100]}$), as a result of $\bf B$-induced deformation of magnetic quadrupoles [see Fig.\ref{fig1}(d)]. Figure \ref{fig4}(a) shows the $T$-dependence of the dielectric constant along the [100] direction ($\varepsilon_{[100]}$) in selected $B_{[100]}$. While no clear anomaly is observed in $B=0$, $\varepsilon_{[100]}$ exhibits a sharp peak at $T_{\rm N}$ in finite $B_{[100]}$, suggesting $B_{[100]}$-induced $P_{[100]}$ due to quadrupole ordering. Indeed, after ME cooling with $B_{[100]}$ and $\bf E$ along the [100] direction ($E_{[100]}$), finite $B_{[100]}$-induced $ P_{[100]}$ emerges below $T_{\rm N}$ on zero-electric-field-warming [Fig.~\ref{fig4}(b)]. Importantly, the sign of $ P_{[100]}$ is reversed by inverting the sign of cooling $E_{[100]}$. Moreover, as seen in Fig. \ref{fig4}(b), $P_{[100]}$ linearly increases with respect to $B_{[100]}$, and this trend is more evident in the approximately straight isothermal $P_{[100]}$-$B_{[100]}$ curve at 2.5 K [Fig.~\ref{fig4}(c)]. These features establish the linear ME activity of the ferroic quadrupole order. The ME coefficient $\alpha_{xx}$ is approximately 8.9 ps/m.
Note that SrTCPO and BaTCPO with the antiferroic quadrupole order exhibit only a much weaker $\varepsilon_{[100]}$ peak in $B_{[100]}$ without accompanying finite $ P_{[100]}$, as seen in Figs.~\ref{fig4}(f) and \ref{fig4}(g) \cite{KKimura2016b}. As discussed in the previous report \cite{KKimura2016b}, this behavior can be interpreted as {\bf B}-induced antiferroelectricity, where {\bf B}-induced {\bf P} in each layer due to the local ME activity cancels out in neighboring layers [Fig.~\ref{fig1}(c)].

We proceed to demonstrate full agreement of the ME activity of PbTCPO with Eq. (4).
First, we have confirmed that except for $\alpha_{xx}$ and $\alpha_{yy}$ all the ME components $\alpha_{ij}$ are zero by observing the absence of $\bf B$-induced $\bf P$ in corresponding measurement configurations \cite{SMprb-PbTi}. Second, we have investigated the effects of $B_{[1\bar{1}0]}$ on $P_{[110]}$ because the relation $\alpha_{yy}=-\alpha_{xx}$ predicts that $B_{[1\bar{1}0]}$ should induce $P_{[110]}$ whose magnitude is the same as that of $B_{[100]}$-induced $ P_{[100]}$. As shown in Figs.~\ref{fig4}(d) and \ref{fig4}(e), this expected $B_{[1\bar{1}0]}$-induced $P_{[110]}$ is clearly observed, with the onset of a sharp $\varepsilon_{[110]}$ peak. Therefore the ME tensor in Eq. (4) is fully valid for PbTCPO.
Due to this highly anisotropic ME activity, the ferroic quadrupole order may induce unique light-polarization-dependent optical phenomena such as nonreciprocal linear dichroism and birefringence \cite{carra2003x}. Exploring these interesting phenomena is left for future work.

The $A$-cation controlled quadrupole order can be reproduced by DFT calculations with GGA+$U$ method \cite{SMprb-PbTi,PBE1996,Liechtenstein1995,Kresse1993,Kresse1994,Kresse1996,Kresse1996_cms,PAW,Kresse1999}. First, following the previous report for BaTCPO \cite{KKimura2016b}, we estimate intralayer magnetic couplings $J_k$ ($k = 1-6$) [Fig.~\ref{fig1}(i)] of SrTCPO and PbTCPO using experimental crystal structures. (Calculations with optimized structures showed insignificant difference \cite{SMprb-PbTi}.) Here, positive (negative) $J$ represents FM (AFM) interactions.
We find that for all the systems, the obtained $J_k$ (Table \ref{tab1}) favor quadrupole order, and their sign and magnitude order are nearly independent of the $A$-cation, except for $J_{5}$.
The $A$-cation-dependent relative strength of $J_{5}$ may be ascribed to the difference in $\phi$ [Fig.~\ref{fig1}(g)], which mostly affects in-plane intercluster couplings like $J_{5}$.
Next, we consider interlayer magnetic couplings by comparing the total energy difference $\Delta E$ between antiferroic and ferroic stacking of magnetic quadrupoles. In agreement with the experimental observation, BaTCPO and SrTCPO favor the antiferroic stacking by 0.10 and 0.22 meV f.u.$^{-1}$, respectively, whereas PbTCPO favors the ferroic stacking by 0.14 meV f.u.$^{-1}$ (Table \ref{tab1}).
Furthermore, we estimate a specific interlayer coupling $J'$ denoted in Fig.~\ref{fig1}(c) because it seems to be directly mediated by the $A$-cation. As shown in Table \ref{tab1}, the values of $J'$ exhibit the same trend upon $A$-cation replacement as $\Delta E$, and the energy difference associated with $J'$, $\Delta E_{J'} = 4J' - (-4J') = 8J'$, is comparable with $\Delta E$. These results suggest that $J'$ is the most dominant interlayer exchange coupling in these systems.

\begin{table}[t]
\centering
\caption{
{Intralayer ($J_{k}$) and interlayer ($J'$) magnetic couplings in meV as well as the energy difference $\Delta E$ in meV f.u.$^{-1}$ between ferroic and antiferroic quadrupole order.}
 \label{tab1}
}
\begin{tabular}{cccccccccccc}
  \hline
  \hline
$A$-ion	 &		$J_1$	&  $J_2$	&	$J_3$	& $J_4$  & $J_5$ & $J_6$ 	&$J'$ & $8J'$ & $\Delta E$ \\
\hline
Ba			&	 -3.03	&	-0.19	&	0.24	&	-0.51	&	-0.65	&	-0.08 		&-0.007	&	-0.052	&	-0.10\\

Sr			&	 -2.95	&	-0.23	&	0.18	&	-0.40	&	-0.25	&	-0.05 		&-0.020	&	-0.162	&	-0.22\\

Pb		&	 -3.00	&	-0.15	&	0.19	&	-0.43	&	-0.42	&	-0.12 		&0.019	&	0.148	&	0.14\\
 \hline
 \hline
\end{tabular}
\end{table}

\begin{figure}[t]
\includegraphics[width=8.6cm]{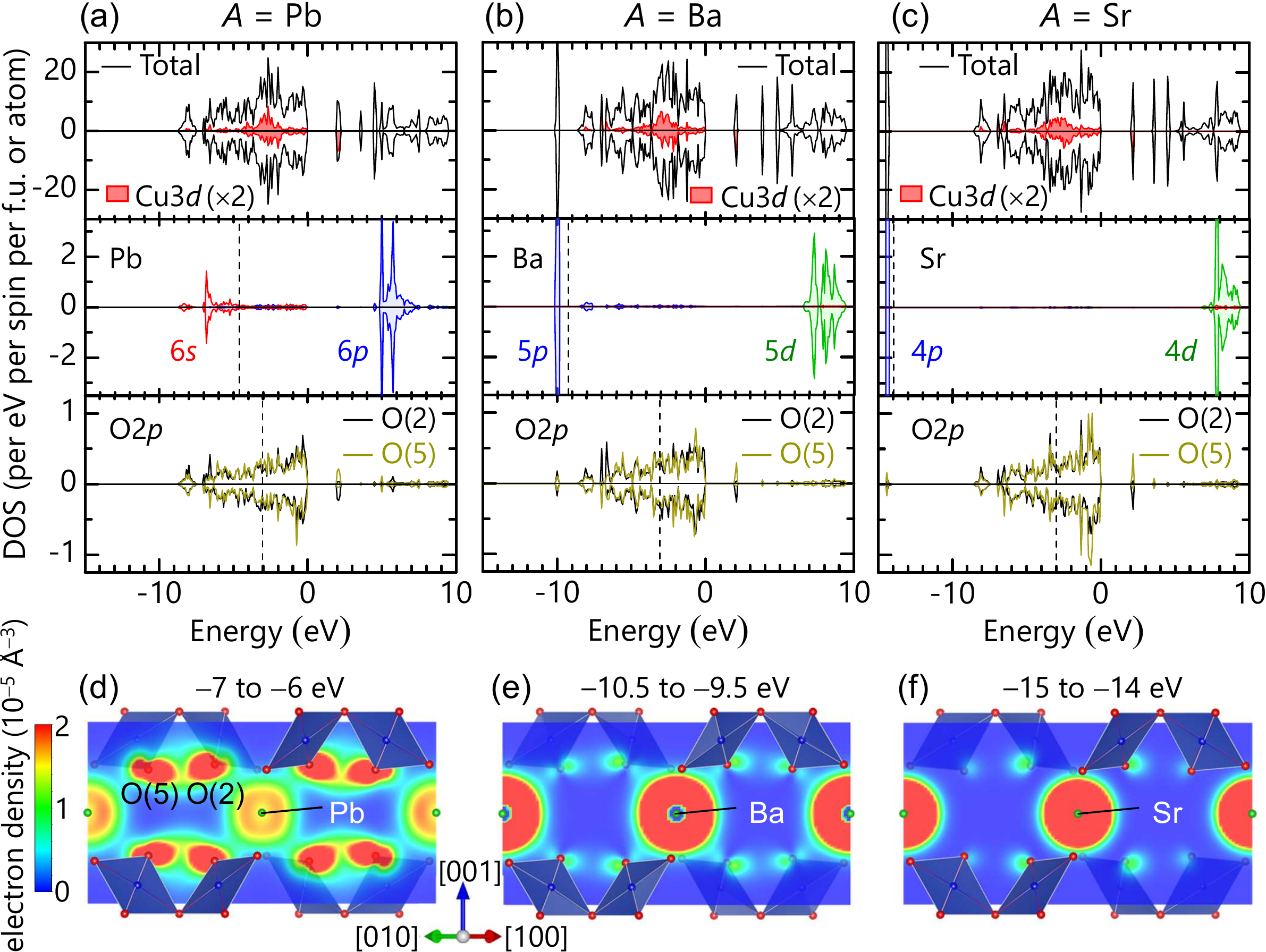}
\caption{
Total DOS and Cu $3d$, $A$-site, and O $2p$ partial DOS of $A$ = (a) Pb, (b) Ba, and (c) Sr systems. The Cu 3$d$ partial DOS is multiplied by 2 for clarity. The dashed vertical lines indicate the band center of mass (see text).
Real-space electron density map for $A$ = (d) Pb, (e) Ba, and (f) Sr systems in the energy window where the $A$ partial DOS is large. The color represents the magnitude of electron density.
\label{fig2}
}
\end{figure}

Because all the structural parameters of PbTCPO are in-between those of SrTCPO and BaTCPO [Figs.~\ref{fig1}(e)-\ref{fig1}(h)], the change of $J'$ cannot be explained by the structural change. We thus attribute it to the electronic states.
Figures \ref{fig2}(a)-\ref{fig2}(c) show total density of states (DOS) and partial DOS of individual elements for the three systems. The O 2$p$ partial DOS is shown only for O(2) and O(5) that are primary associated with $J'$ [Figs.~\ref{fig1}(a) and \ref{fig1}(c)]. The basic feature of the total DOS is cation-independent and the top of valence band and the bottom of conduction band are formed by hybridized Cu 3$d$ and O 2$p$ orbitals. The distinct difference is found in the $A$ partial DOS.
This affects the degree of $A$-O orbital hybridization, as visualized in Figs.~\ref{fig2}(d)-\ref{fig2}(f). Strong $A$-O hybridization is seen in PbTCPO and it clearly bridges the Cu$_4$O$_{12}$ clusters of neighboring layers. This suggests that the strong $A$-O hybridization is responsible for FM $J'$.

For more  discussion, we quantify the degree of $A$-O hybridization using a bond covalency defined as $C_{{\rm A,O}} = -|C_{\rm MA}-C_{\rm MO}|$ (in eV) \cite{cammarata2014microscopic,cammarata2014covalent}.
$|C_{\rm MA}|$ and $|C_{\rm MO}|$ are the band center of mass for $A$ orbital (Pb 6$s$, Ba 5$p$, and Sr 4$p$) and O 2$p$ orbital averaged over O(2) and O(5), respectively, obtained through a $k$-space integration over an energy window from  $-15$ to 0 eV.
Dashed vertical lines in Figs.~\ref{fig2}(a)-\ref{fig2}(c) denote the calculated $|C_{\rm MA}|$ and $|C_{\rm MO}|$, which yield the covalency $-1.57$ (Pb) $ > -6.17 $ (Ba) $>-10.97$ (Sr). This means the strongest $A$-O hybridization for PbTCPO. Comparing this result with $J'$ in Table \ref{tab1}, we find that ferroic quadrupole order becomes more stable for a more covalent $A$-O bond.

Finally, we predict a new ferroic quadrupole system based on the obtained covalency-interlayer coupling relation. A good candidate is Sn(TiO)Cu$_4$(PO$_4$)$_4$ because Sn$^{2+}$  is one of $s^{2}$-cations that often form a strong covalent bond with O ions. Indeed, DFT calculations show the strong Sn-O covalency of -0.32 eV and FM $J'$, which predicts that ferroic quadrupole order is realized in Sn(TiO)Cu$_4$(PO$_4$)$_4$ \cite{SMprb-PbTi}.

\section{CONCLUSION}
In summary, our experiments on $A$(TiO)Cu$_4$(PO$_4$)$_4$ ($A$ = Ba, Sr, and Pb) show that inserting stereochemically inactive $s^2$-cation Pb$^{2+}$ into $A$-site switches the arrangement of magnetic quadrupoles from antiferroic to ferroic, resulting from the sign reversal of the specific interlayer magnetic coupling. DFT calculations elucidate that the sign of the interlayer coupling is correlated with the $A$-O bond covalency and the ferromagnetic coupling is realized when the covalency is strong, as in the $A$ = Pb system.
The present result opens up a strategy for designing magnetism by tuning bond covalency, where an $s^2$-cation which often forms a strong covalent bond can be a good tuning option. This stimulates the search for various exotic magnetism such as magnetoelectric order {\cite{Spaldin2008} and quantum spin liquid \cite{Balents2010}, which may be applied to magnetoelectric memory and new optical and spintronic devices.
\\\
\\\

\begin{acknowledgments}
We wish to thank Y. Kato and Y. Motome for helpful discussions, H. Tada for specific heat measurements, and T. Hansen and C. Ritter for their assistance during the D20 experiment.
This work was partially supported by JSPS KAKENHI Grant Numbers JP26610103, JP16K05449, JP26800186, and JP17H01143, by the MEXT Leading Initiative for Excellent Young Researchers (LEADER), by a research grant from The Murata Science Foundation, by European Research Council grant CONQUEST, and by the Swiss National Science Foundation and its Sinergia network Mott Physics Beyond the Heisenberg Model (MPBH).
\end{acknowledgments}



%

\end{document}